\documentclass[showpacs,preprintnumbers,preprint,aps,amsmath,amssymb]{revtex4}
\usepackage{epsfig}
\usepackage{graphics}
\def\lsim{\:\raisebox{-1.1ex}{$\stackrel{\textstyle<}{\sim}$}\:}
\def\gsim{\:\raisebox{-1.1ex}{$\stackrel{\textstyle>}{\sim}$}\:}
\def\10{$SO(10)$}
\def\bea{\begin{eqnarray}}
\def\eea{\end{eqnarray}}
\def\21{SU(2) $\otimes$ U(1) }

\def\422{$SU(4) \otimes SU(2) \otimes SU(2)$}
\def\321{SU(3) $\otimes$ SU(2) $\otimes$ U(1)}
\def\lsim{\raise0.3ex\hbox{$\;<$\kern-0.75em\raise-1.1ex\hbox{$\sim\;$}}}
\def\gsim{\raise0.3ex\hbox{$\;>$\kern-0.75em\raise-1.1ex\hbox{$\sim\;$}}}
\def\vev#1{\left\langle #1\right\rangle}

\newcommand{\ba}{\begin{array}}
\newcommand{\ea}{\end{array}}
\newcommand{\be}{\begin{equation}}
\newcommand{\ee}{\end{equation}}
\newcommand{\beqa}{\begin{eqnarray}}
\newcommand{\eeqa}{\end{eqnarray}}
\relax
\def\321{$SU(3)\times SU(2)\times U(1)$}

\newcommand{\bsbsbar}{$B^0_s-\bar{B}^0_s~~$}

\begin{document}
\bigskip
\title{Higgs induced  FCNC as a source of new physics in $b\rightarrow s $ transitions} 
\bigskip
\author{ Anjan S.  Joshipura\footnote{anjan@prl.res.in} and  Bhavik P.  
Kodrani\footnote{bhavik@prl.res.in} } \affiliation{ Physical Research
Laboratory, Navarangpura, Ahmedabad 380 009, India \vskip 2.0truecm}

\begin{abstract} 
\vskip 1.0 truecm 
The observations in the   $B_{s}$ sector suggest the existence of  some new physics contribution to the \bsbsbar mixing.
We study the implications of a hypothesis that this contribution is generated by the 
Higgs induced flavour changing neutral currents. We concentrate on the specific $b\rightarrow s$ transition which is
described by two complex FCNC parameters  $F_{23}$ and $F_{32}$ and  parameters in the Higgs sector. 
Model-independent constraints on these parameters are derived from the \bsbsbar mixing and are used 
to predict the branching ratios
for $\bar{B}_s\rightarrow \mu^+\mu^-$ and $\bar{B}_d\rightarrow \bar{K}\mu^+\mu^-$ numerically by considering general variations in the Higgs parameters assuming that Higgs sector conserves CP.   Taking the results on  $B_{s}^0-\bar{B}_{s}^0$ mixing derived by the global analysis of UTfit group as a guide
we present the  general constraints on $F_{23}^*F_{32}$ in terms of the pseudo-scalar mass $M_A$. The former is required to be in the 
range  $\sim (1-5) \times 10^{-11} M_A^2 {\rm GeV}^{-2}$ if the Higgs induced FCNC represent the dominant source of new physics. The
phases of these couplings can account for the large CP violating phase in   the $B_{s}^0-\bar{B}_{s}^0$ mixing except when $F_{23}=F_{32}$.
The Higgs contribution to $\bar{B}_s\rightarrow \mu^+\mu^-$ branching ratio can be large, close to the present limit while it remains close to the standard model value in case of the process 
$\bar{B}_d\rightarrow \bar{K}\mu^+\mu^-$  for all the models under study. 
We identify and discuss various specific examples which can naturally lead to suppressed FCNC in the $K^0-\bar{K}^0$ mixing allowing at the same time the required values for $F_{23}^*$ and $F_{32}$.   
\end{abstract} 

\maketitle
\newpage 
\section{Introduction}
The Cabibbo Kobayashi Maskawa (CKM) matrix $V$ provides a unique
source of flavour and CP violations in the standard model (SM). It leads
to flavour changing neutral currents (FCNC) at the one loop level.
$K$ and $B$ meson decays and mixing have provided stringent
tests of these FCNC induced processes and the SM predictions have been 
verified  with  some hints for 
possible new physics contributions \cite{np1,bf,np2}. Any 
new source 
of flavour violations 
resulting from the well-motivated extensions of the SM ($e.g.$ 
supersymmetry) is now constrained  to be small \cite{utfit,ckmfit}.

Uncovering highly constrained new physics becomes easier if one  specifically looks at
observables which are predicted to be small or zero in the SM. Transitions between the $b$ and $s$ quarks
offer such observables \cite{lenz}.  The $b\leftrightarrow s$ transitions among other things lead to (1) $\Delta B=2$, \bsbsbar  mixing (2) the leptonic decays $\bar{B}_s\rightarrow
l^+l^- ~(l=e,\mu,\tau)$ (3) The semi leptonic decays $\bar{B}_d\rightarrow (\bar{K},\bar{K}^*)\mu^+\mu^-$ . The CP violating phase 
$$ \phi_s=Arg(-\frac{M_{12}}{\Gamma_{12}})$$ where $M_{12}$ and $\Gamma_{12}$ respectively denote the real and absorptive parts of the \bsbsbar  transition amplitude is predicted to be quite small $\sim 0.2^{\circ}$ in the SM .  In contrast, the experimental determination of $\phi_s$
from the time-dependent CP asymmetry in $B_s\rightarrow J/\psi\phi$ decays by the D0 \cite{d0} and CDF \cite{cdf}  groups allow much larger phase:
the 90\% CL average reported by HFAG \cite{hfag}
requires $[-1,47;-0.29]\cup[-2.85;-1.56]$. By including the D0 and the CDF results in their global analysis, UTfit 
group find around $3\sigma$ departure from the SM prediction on $\phi_s$ \cite{utfit, Bonna}. Similar analysis by the CKMfitter group
\cite{ckmfit}  also reports deviation from the SM result but at around $2.5 \sigma$. This may be a  hint of the presence of new physics in the $b\leftrightarrow s$ transitions. Future measurement would provide a crucial test of this possibility.

The decay rate for $\bar{B}_s\rightarrow \mu^+\mu^-$ is also predicted to be small in SM
\be \label{bsmumusm} Br(\bar{B}_s\rightarrow\mu^+\mu^-)=(3.51\pm0.50)\times 10^{-9}~. \ee
compared to an order of magnitude larger experimental limit
\be \label{bsmummu} Br(\bar{B}_s\rightarrow\mu^+\mu^-)<5.8\times 10^{-8}~~~{\rm ( 95\% CL) } ~. \ee
This rate therefore can be an important observable  in  search of new physics.
In contrast, the branching ratios for the exclusive processes in (3) are close to the SM predictions. But 
they still provide valuable constraints on any new physics that may be present.  Moreover, the di-lepton spectrum and the angular distribution of leptons in these exclusive 
processes provide very sensitive test of the SM and possible indication of new physics \cite{bobeth1,bobeth2}. The LHCb \cite{lhcb} and the super-B factory will allow more sensitive determination of these observables and will strongly constrain or uncover any new physics that may be present. 
 
The $b\leftrightarrow s$ transition is also interesting from the theoretical point of view since several extensions of SM 
predict relatively large effects in this transition.  The most popular extensions studied are the two Higgs doublet models (2HDM)  in which some symmetry (discrete or super) prevents FCNC at the tree level.  In these models, the Higgs (like the W boson) contribute to the FCNC at the loop level. The supersymmetric standard model is one such example within which  the Higgs and sparticle mediated  flavour changing effects have been extensively studied \cite{mfv1}.
In the  Minimal Supersymmetric Standard Model (MSSM),
the $d_i\leftrightarrow d_j$ transitions between the charged $-1/3$ quarks in large $\tan\beta$ limit are governed by the CKM factor $V_{3i}V^*_{3j}$ \cite{babukolda,buras}. As a result, the effect becomes more prominent for the $b\leftrightarrow s$ transitions compared to others. The same thing also happens in the charged Higgs induced flavour transitions in the two Higgs doublet model with the natural flavour conservation (NFC) .  Both these cases realize the Minimal Flavour Violation (MFV) \cite{mfv2} and do not have any additional CP violating phase other than the CKM phase. In the context of the MSSM, one can consider scenarios which go beyond the MFV to accommodate a large $\phi_s$ \cite{mfv1,beyondmfv}. This cannot easily be done for two Higgs doublet model with NFC. Large CP violating phases are possible in more general two Higgs doublet models ( called type - III 2HDM ) which allow the tree level FCNC. Most general model of this type can lead to large flavour violation in the $d\leftrightarrow s$ transitions and would imply a very heavy Higgs mass suppressing all other flavour violations. It is possible to imagine  scenarios where the tree level FCNC couplings also show hierarchy as in the quark masses \cite{hall}. This class of models would imply relatively large flavour violations in $B$ transitions. The standard example of this is the so called
Cheng- Sher ansatz \cite{chengsher}  which postulates a relation between the down quark masses $m_i$ and the FCNC couplings:
\be \label{chengsher} F_{ij}= \lambda_{ij}\frac{\sqrt{m_im_j}}{v} ~, \ee
with $\lambda_{ij}\sim {\cal O}(1)$ and $v\sim O(174{\rm GeV})$. 

There exist explicit models \cite{asj1,bgl,ars} which lead to
hierarchy in FCNC. Such models which are theoretically as natural as the two Higgs doublets with NFC can lead to interesting
patterns of flavour violations. Our aim in this paper is to analyze the constrains and prediction of the Higgs induced tree level FCNC in the $b\leftrightarrow s$ transitions. Rather than looking at any specific model in this category we consider several classes of models which imply interesting patterns of flavour violation. We find that the predictions of some of these models for the leptonic and semi leptonic transitions mentioned above are distinctively different compared to the two Higgs doublet  models with NFC and the MSSM.
Moreover, it is possible within them to simultaneously look at the constraints from all three processes listed above
and we find that the \bsbsbar  mixing provides very stringent restrictions on the other two processes.

There have been earlier phenomenological studies of models with tree level FCNC \cite{type3phen}. Most of these are model specific
and mainly use the Cheng-Sher ansatz and try to constrain parameters $\lambda_{ij}$. As we discuss, there are models which are distinctively different from this ansatz. So rather than specifying any specific model, we perform a model-independent analysis
of the Higgs induced FCNC couplings. Unlike the Cheng-Sher ansatz, these couplings in general can have phases which are not included in the earlier analysis. As we show, the FCNC couplings may provide the source of a large $\phi_s$ and we identify  models which explain large $\phi_s$ and those which can not do so.

We present the general structure of the Higgs induced FCNC in the next section where we also discuss various classes of models
which lead to hierarchical FCNC couplings. In section (III), we give the details of the effective Hamiltonian for the $\Delta B=1$ and $2$ transitions. In the next section, we derive an important relation between the Higgs contributions to the  \bsbsbar mass difference and to the branching ratio for $\bar{B}_s\rightarrow\mu^+\mu^-$. This relation is independent of the FCNC couplings $F_{23}^*,F_{32}$ under specific assumptions. In the same section, we study numerical implications of various classes of models and conclude in the last section.
\section{FCNC: Structure and examples}
This section is devoted to a discussion of classes of the 2HDM which we use as a guide to carry out  a fairly model-independent analysis of the $b\rightarrow s$ transitions subsequently.

The general two Higgs doublet models \cite{hk}   have the following Yukawa couplings in the down quark sector:
\be \label{yukawa}
-{\cal L}_Y^d= \bar{d}_L' (\Gamma_1 \phi_1^0+\Gamma_2 \phi_2^0)d_R'+{\rm H.c.} ~. \ee
Here, $d_{L,R}'$ denote (the column of) the weak eigenstates of down quarks.  The models with NFC impose an additional discrete symmetry, {\it e.g.}  $(d_R',\phi_1)\rightarrow -(d_R',\phi_1)$ which forbids the couplings $\Gamma_2$. As a result, the down quark couplings to $\phi_1$ become diagonal in the mass basis and there are no tree level FCNC.  

More general 2HDM allow both  $\Gamma_{1}$ and $\Gamma_2$ in eq.(\ref{yukawa}) and contain the tree level FCNC. 
Consider  two orthogonal combinations of the Higgs fields $\phi_{1},\phi_2$:
\bea \label{fchiggs}
\phi^0&\equiv& \cos\beta \phi_1^0+\sin\beta \phi_2^0 ~, \nonumber\\
\phi^0_F&\equiv& -\sin\beta \phi_1^0+\cos\beta \phi_2^0 ~,\\
\eea
with $\vev\phi_1=v \cos\beta~;~ \vev\phi_2=v \sin\beta$ and $v\sim 174$ GeV. $\phi^0$ acquires a non-zero vacuum expectation value (vev)  and leads to the quark mass matrix
\be \label{Md}
M_d=v(\Gamma_1 \cos\beta+\Gamma_2 \sin\beta) ~.\ee
$\phi^0$ is like the SM Higgs  field with flavour conserving couplings to quarks. The $\phi_F^0$  violates flavour and one can write using, eqs.(\ref{yukawa},\ref{Md})  
\be \label{fcncint}-{\cal L}_{FCNC}=\sum_{i\not =j} F_{ij} \bar{d}_{iL} d_{jR}\phi_F^0+{\rm H.c.} ~.\ee
$d_{L,R}$ denote the mass eigenstates, 
\be \label{fcnc}
F_{ij}\equiv (V_L^\dagger \Gamma_2 V_R)_{ij}\frac{1}{\cos\beta} ~. \ee
and $V_{L,R}$ are defined by 
\be \label{diag} V_L^\dagger M_d V_R=D_d ~.\ee
Here $D_d$ is the diagonal mass matrix for the down quarks. The structure as in (\ref{fcncint}) can arise as an effective interactions from the loop diagrams as in MSSM \cite{babukolda} or the 2HDM with NFC\cite{isidori1}. Phenomenology based on this structure therefore would include such cases also.

The leptonic and semi-leptonic FCNC transitions also depend on how the charged leptons couple to the fields $\phi_{1,2}$.
For definiteness, we will assume that the charged lepton Yukawa couplings are given as in the MSSM.  We thus assume

\beqa \label{yukawal}
-{\cal L}_Y^l&=& \bar{l}_{L}' \Gamma^l_1 l_R'\phi_1^0+{\rm H.c.} ~,\nonumber\\
&=&\frac{1}{v\cos\beta}\bar{l}_{L}D_l l_R\phi_1^0+{\rm H.c.} ~. \eeqa
If coupling to $\phi_2$ is also present then one would get flavour violations in the leptonic sector also.

General properties of $F$ follow from its definition, eq.(\ref{fcnc}). We shall consider three specific class of FCNC and show that each of these imply different and interesting physics.\\ 
\noindent (A) Hermitian structures: Assume that quark mass matrices and $\Gamma_{1,2}$ are Hermitian. In this case, eq.(\ref{fcnc}) trivially implies
\be \label{hermitian}  F_{ij}=F_{ji}^*~. \ee\\
\noindent  (B) Symmetric structures: Assume that $M_d$ and $\Gamma_{1,2}$ are symmetric. This trivially leads to symmetric FCNC couplings:
\be \label{symmetric}  F_{ij}=F_{ji}~. \ee\\
\noindent  (C) MSSM like structures:
The FCNC in  MSSM in  large $\tan\beta$ limit \cite{babukolda, buras} can be described by an effective tree level Lagrangian similar to eq.(\ref{fcnc}) with the specific relation
\be \label{mssmlike}  F_{ij}=\frac{m_j}{m_i} F_{ji}^* ~ \ee
between the FCNC couplings. The same relation also holds in general 2HDM with NFC where $F_{ij}$ are induced by the 
charged Higgs at 1-loop \cite{isidori1}. More interestingly, even the tree level FCNC can satisfy the same relation in some specific models \cite{asj1,bgl}. 

While the phenomenological analysis that we present in the above three cases would be model independent,  we give below several examples of textures/models which can realize above scenarios and simultaneously explain the quark masses. \\ \\
\noindent {\bf Yukawa textures and FCNC}\\ \\
The strongest constraints on FCNC come from the $K^0-\bar{K}^0$ mixing and the $\epsilon$ parameter. One needs very heavy Higgs$\sim$ O(TeV)  to suppress this effect if $F_{12}\sim $ O(gauge coupling). Heavy Higgs would then suppress other flavour violations as well  without leaving any signature at low energy.  Interesting class of models would be the ones in which the coupling $|F_{12}|$ would be suppressed compared to the other couplings. As already discussed in the introduction, widely studied example of this is the Cheng-Sher ansatz, eq.(\ref{chengsher}).  Here the suppression in $F_{ij}$ comes from the suppression in the quark masses compared to the weak scale.  $F_{ij}$ may also be suppressed by mixing angles.  This can come about naturally in large classes of 2HDM.  Assume  that  the Higgs $\phi_2$ in eq.(\ref{yukawa}) is responsible for only the third generation mass while the Higgs $\phi_1$ accounts for the first two generation masses and the inter-generation mixing. Only the (33) element of  $\Gamma_2$ is assumed non-zero in this case and eq.(\ref{fcnc}) automatically implies
\be \label{fcnc1} F_{ij}=\frac{m_b}{v\cos \beta\sin\beta} V_{L3i}^* V_{R3j} ~. \ee
If $M_d$ is Hermitian or symmetric one automatically obtains eq.(\ref{hermitian}) or (\ref{symmetric}). If  the off-diagonal elements of $V_{L,R}$ are suppressed compared to the diagonal elements, then  $F_{12}$ will be more suppressed compared to others. In particular,  $(V_{L,R})_{ij}\sim c_{L,R} \sqrt{\frac{m_i}{m_J}}, (i < j)$  reproduces the Cheng-Sher ansatz with $\lambda_{ij}\sim \frac{c_Lc_R}{\cos\beta\sin\beta}$.  Thus this class of models may be regarded as a generalization of the Cheng Sher ansatz.

Let us take two concrete examples which are among the specific textures studied in the literature with a view to  understand the fermion masses and mixings.

Consider
\begin{itemize}
\item
\begin{equation}
\Gamma_1=y_{33}\left( \ba{ccc} d \epsilon^4 & b\epsilon^3& c\epsilon^3\\
                                                     b \epsilon^3& f \epsilon^2& a \epsilon^2\\
                                                     c \epsilon^3&a \epsilon^2& 0\\ \ea \right) ~;~
\Gamma_2=y_{33}\left( \ba{ccc}       0&0&0\\
                                                     0& 0&0\\
                                                     0&0&1\\ \ea \right)  ~.  \ee
These couplings together imply the down quark mass matrix studied long ago by Roberts, Romanino, Ross and Velesco-Sevilla \cite{RRRS} and  recently in\cite{ross}. 
 $\epsilon$ here is a small parameter which can be determined from the quark masses.  $\epsilon\sim 0.1$ is determined in \cite{ross} assuming the above structure to be valid at the GUT scale.
Above matrices imply in a straight forward way
\be
|V_{L32}|=|V^*_{R32}|\sim a \epsilon^2~; |V_{L31}|=|V^*_{R31}|\sim |c| \epsilon^3~;|V_{L12}|=|V^*_{R12}|\sim \frac{b}{f} \epsilon ~. \ee
This in turn implies
\be
|F_{12}|\approx \frac{m_b}{v\cos\beta \sin\beta} a |c| \epsilon^5~; |F_{13}|\approx \frac{m_b}{v\cos\beta \sin\beta} |c| \epsilon^3~; |F_{23}|\approx \frac{m_b}{v\cos\beta \sin\beta}a\epsilon^2 ~.\ee
Thus one obtains the desired hierarchical FCNC couplings with this ansatz.
\item
As an other example we consider the texture suggested in \cite{ds}:
\begin{equation}
\Gamma_1=y_{33}\left( \ba{ccc} d \epsilon^6&b \epsilon^4&c \epsilon^3\\
                                                     b \epsilon^4& f \epsilon^2& a \epsilon\\
                                                     c \epsilon^3& a \epsilon&0\\ \ea \right) ~;~
\Gamma_2=y_{33}\left( \ba{ccc}       0&0&0\\
                                                     0& 0&0\\
                                                     0&0&1\\ \ea \right)  ~.  \ee
 where $\epsilon$ is a small expansion parameter   (assumed to be $\sim 0.2$ in \cite{ds}) and other parameters are O(1).
The quark mass matrix is of rank 1 if these parameters are exactly 1. Because of this feature, it is possible to simultaneously understand the large mixing in the neutrinos and small mixing in the quark sector. The above form of the quark matrix also implies the relation $$ (V_L)_{ij}\approx \sqrt{\frac{m_i}{m_j}}~, (i < j).$$
As a result, the FCNC couplings satisfy the Cheng-Sher ansatz given in eq.(\ref{chengsher}) with $\lambda_{ij}\sim \frac{1}{\cos\beta \sin\beta}$. 
$M_d$ and  Yukawa couplings are symmetric in both the above examples. One could consider instead similar Hermitian textures as well. \\
\item Somewhat different illustration of the suppressed FCNC couplings is provided by the following textures  of the Yukawa couplings:
\begin{equation}
\Gamma_1=\left( \ba{ccc} x&x&x\\
                                                     x& x& x \\
                                                     0& 0&0\\ \ea \right) ~;~
\Gamma_2=\left( \ba{ccc}       0&0&0\\
                                                     0& 0&0\\
                                                     x&x&x\\ \ea \right)  ~,\ee
where $x$ denotes an entry which is not required to be zero. It is straightforward to show that the above Yukawa couplings imply
\be \label{mfv} F_{ij}=\frac{1}{v\cos\beta \sin\beta}V_{L3i}^* V_{L3j} m_j~\ee
and therefore satisfy relation (\ref{mssmlike}).
Note that $F_{ij}$ depend only on the left-handed mixing matrix and they remain suppressed and hierarchical if the mixing elements show hierarchy.  The structure of FCNC in this example is different compared to the Cheng-Sher ansatz and earlier two examples.  
 The earlier two examples reduce to the Cheng-Sher ansatz if $V_{Lij}\approx \sqrt{\frac{m_i}{m_j}}, (i < j) $ while eq.(\ref{mfv}) has an additional suppression by
$\frac{m_j}{m_b}$ compared to them in this case when $j\not=3$. 

This particular example
of the suppressed FCNC couplings was  proposed in \cite{asj1}.
The hierarchy among $F_{ij}$ is determined in the MSSM by the CKM matrix elements while here it is determined by the elements of the down quark mixing matrix. In particular, the $F_{ij}$ can have new phases not present in the MSSM case. It is possible to construct models \cite{bgl} in which  $V_L$ in eq.(\ref{mfv}) gets replaced by the CKM matrix making the $F_{ij}$ very similar to the MSSM model. Phenomenological consequences of this were studied in \cite{bhavik1}.
\end{itemize}
The examples given here are representative rather than exhaustive. One could consider several similar structures, e.g.
one based on the Fritzsch ansatz \cite{textures} or on some different textures  , e.g. based on the $\mu$-$\tau$ interchange symmetry \cite{bhavik2} 
all with the property of the suppressed and hierarchical FCNC. Without subscribing to any specific model we shall now
consider the general implications for the $b\leftrightarrow s$ transitions.
\section{Effective Hamiltonian for the $b\leftrightarrow s$ transitions}
The basic interaction in eq.(\ref{fcncint}) leads to both  $\Delta B=1$ and $2$ transitions. We give below the corresponding effective Hamiltonian.
\subsection{\bsbsbar mixing}
\bsbsbar mixing is governed by the transition amplitude \cite{buraslectures}
$$ M_{12}^{*s}\equiv \langle \bar{B}^0_s|{\cal H}_{eff}|B_s^0\rangle ~. $$
Here, $${\cal H}_{eff}\equiv {\cal H}_{eff}^{SM}+{\cal H}_{eff}^{NP}$$ includes the SM and the new physics contribution to the
\bsbsbar transition. The $\phi_F^0$ exchange leads to three new operators contributing to the \bsbsbar transition:
$$ Q_2^{LR}=(\bar{b}_Ls_R)(\bar{b}_Rs_L)~~; Q_1^{LL}=(\bar{b}_Rs_L)(\bar{b}_Rs_L)~~;Q_3^{RR}=(\bar{b}_Ls_R)(\bar{b}_Ls_R)$$
Taking the matrix elements of these operators between the $\bar{B}^0_s$ and $B^0_s$  mesons in ${\cal H}^{NP}$  one arrives at \cite{buraslectures}
\be \label{m12s} 
(M_{12}^{s*})^{NP}=\frac{G_F^2M_W^2}{48 \pi^2} M_{B_s}f_{B_s}^2 (V_{tb}^*V_{ts})^2 \left[ P_2 C_2^{LR} +P_1 C_1^{LL}+P_1 C_3^{RR}\right] ~. \ee
Here, $G_F$ is the Fermi coupling and $M_{B_S},f_{B_s}$ are the mass and the decay constant of the $B_s$ meson.
$V$ denotes the CKM matrix. The $C_{1,2}$ above refer to the Wilson coefficients evaluated at the Higgs mass scale. $P_{1,2}$ summarize the effect of the evolution to the low scale and the Bag factors. When Higgs scale is identified with the top mass one gets, $P_2\approx 2.56$ and $P_1\approx -1.06$ \cite{buraslectures}. For definiteness, we will use these values in the numerical analysis. The Wilson
coefficients are given in our case as
\beqa \label{c12}
C_{2}^{LR}&=&-\frac{ 16 \pi^2}{ G_F^2M_W^2 (V_{tb}^*V_{ts})^2} F_{32} F_{23}^* \langle\phi_F|\phi_F^*\rangle~,\nonumber\\
C_{1}^{LL}&=&-\frac{1}{2}\frac{ 16 \pi^2}{ G_F^2M_W^2 (V_{tb}^*V_{ts})^2} F_{23}^{*~2} \langle\phi_F^*|\phi_F^*\rangle, \nonumber \\
C_{3}^{RR}&=&-\frac{1}{2}\frac{ 16 \pi^2}{ G_F^2M_W^2 (V_{tb}^*V_{ts})^2} F_{32}^2 \langle\phi_F|\phi_F\rangle~.
 \eeqa
Here $\langle \phi_F|\phi_F\rangle$ etc are proportional to various  propagators and are defined below. The total mixing amplitude is given by
\be \label{kapasdef}
M_{12}^{s*}=(M_{12}^{s*})^{SM}+(M_{12}^{s*})^{NP}\equiv (M_{12}^{s*})^{SM}(1+\kappa_s^H e^{2 i (\phi_s^H+\beta_s)})~, \ee

The $(M_{12}^{*s})^{SM}$ is given \cite{buraslectures} by
\be\label{m12sm}
(M_{12}^{*s})^{SM}=\frac{G_F^2 M_W^2 M_{B_s}f_{B_s}^2 B_{B_{s}}\eta_B}{12 \pi^2}
(V_{tb}^* V_{ts})^2 S_0(x_t) ~, \ee
with $S_0(x_t)\approx 2.3$ for $m_t\approx 161$ GeV and $\eta_B\approx 0.55$ 
represents the QCD corrections.
Using eq.(\ref{c12}) we find
\be \label{kapas}
\kappa_s^H e^{2 i \phi_s^H}=-\frac{4 \pi^2 }{B_{B_s}\eta_B S_0(x_t)G_F^2M_W^2 |V_{tb}^*V_{ts}|^2} \left[ P_2 F_{32} F_{23}^* \langle\phi_F|\phi_F^*\rangle +\frac{1}{2}P_1(F_{32}^2 \langle\phi_F|\phi_F\rangle+F_{23}^{*2} \langle\phi_F|\phi_F\rangle^*)\right] ~. \ee

The new physics induced phase in the above expression is determined by the phases of the FCNC couplings and the complex
Higgs propagators. We assume throughout that the Higgs sector is CP conserving. In this case, the only source of the non-standard CP violation resides in the phases of $F_{23},F_{32}$. Various propagators can be written under the above assumption in terms of the mass eigenstates fields, the scalars $h,H$ and  a pseudoscalar $A$ given by
\beqa \label{higgs}
h&=&\sqrt{2}(\cos\alpha Re(\phi_2)-\sin\alpha Re(\phi_1)) ~,\nonumber \\ 
H&=&\sqrt{2}(\sin\alpha Re(\phi_2)+\cos\alpha Re(\phi_1)) ~,\nonumber \\ 
A&=&\sqrt{2} Im(\phi_F) ~. \eeqa
The propagators appearing in eq.(\ref{kapas}) are now given by
\begin{eqnarray} \label{prop}
\langle \phi_F| \phi_F^*\rangle &=& \frac{\text{sin}^2(\alpha - \beta)}{2 M_H^2} +  \frac{\text{cos}^2(\alpha - \beta)}{2 M_h^2} + \frac{1}{2 M_A^2} ~, \nonumber\\
\langle\phi_F |\phi_F\rangle &=& \langle \phi_F| \phi_F \rangle^* = \frac{\text{sin}^2(\alpha - \beta)}{2 M_H^2} +  \frac{\text{cos}^2(\alpha - \beta)}{2 M_h^2} - \frac{1}{2 M_A^2}~.
\end{eqnarray}
\subsection{$\Delta B=1$ transitions}
The transition $b\rightarrow s$ occurs in SM at the 1-loop level.  The corresponding effective Hamiltonian is described in terms of 
10 different operators and associated Wilson coefficients.  The complete list  can be found for example in \cite{bobeth1}. The Wilson coefficients are calculated at the electroweak scale and are then evaluated in the low energy theory in a standard way. If some new physics is present at or above the electroweak scale then (1) it can give additional contributions to some of the  Wilson coefficients  and/or (2) can lead to new sets of operators not present in the SM.  We will mainly be concerned here with  effects due to  (2) induced by the presence of the  non-standard Higgs field(s) but the effect (1) may also be simultaneously present.

The Higgs induced operators  for the transition  $b\rightarrow s \mu^+ \mu^-$ may be parametrized as: 
\be \label{ope}  {\cal H}_{eff}^H\equiv- \frac{4 G_F}{\sqrt{2}} V_{tb} V_{ts}^*\sum_{i=S,S',P,P'}  C_i(\mu)O_i(\mu) ~. \ee
where $\mu$ denotes the 
renormalization scale at which the operators and the Wilson coefficients appearing above are defined. The operators are defined as
\beqa  \label{opdef}
O_S=\frac{e^2}{16 \pi^2}\bar{s}_L b_R \bar{\mu}\mu&;& O_P= \frac{e^2}{16 \pi^2}\bar{s}_L b_R \bar{\mu}\gamma_5\mu ~\nonumber \\
O_S'=\frac{e^2}{16 \pi^2}\bar{s}_Rb_L \bar{\mu}\mu&;& O_P'=\frac{e^2}{16 \pi^2} \bar{s}_R b_L \bar{\mu}\gamma_5\mu ~, \eeqa
The tree level Higgs exchange through eq.(\ref{fcncint})   induce the above operators with the Wilson coefficients given by
\begin{eqnarray}\label{cscp}
C_S &=& -\frac{\sqrt{2} \pi}{\alpha G_F V_{tb}V_{ts}^*}\frac{F_{23} m_\mu}{2 v \cos \beta} \Big(\frac{\sin(\alpha - \beta) \cos \alpha}{M_H^2} - \frac{\cos(\alpha - \beta) \sin \alpha}{M_h^2}\Big) ~, \nonumber\\ 
C_S' &=&  -\frac{\sqrt{2} \pi}{\alpha G_F V_{tb}V_{ts}^*}\frac{F_{32}^* m_\mu}{2 v \cos \beta} \Big(\frac{\sin(\alpha - \beta) \cos \alpha}{M_H^2} - \frac{\cos(\alpha - \beta) \sin \alpha}{M_h^2}\Big)  ~,\nonumber\\
C_P &=& - \frac{\sqrt{2} \pi}{\alpha G_F V_{tb}V_{ts}^*} \frac{F_{23} m_\mu}{2 v \cos \beta} \frac{\sin \beta}{M_A^2} ~, \nonumber\\
C_P' &=& -\frac{\sqrt{2} \pi}{\alpha G_F V_{tb}V_{ts}^*}\frac{F_{32}^* m_\mu}{2 v \cos \beta}  \Big(-\frac{\sin \beta}{M_A^2}\Big)~.
\end{eqnarray}

Eq.(\ref{ope}) contributes both  to the $\bar{B}_s\rightarrow \mu^+\mu^-$ and the $\bar{B}_d\rightarrow \bar{K}(\bar{K}^*) \mu^+\mu^-$ processes. The Higgs contribution to the branching ratio for the former process follows \cite{bobeth1, isidori1} in a straightforward way from eq.(\ref{ope}):
\begin{eqnarray} \label{bsh}
Br(\bar{B}_s\rightarrow \mu^+\mu^-)&=&\left( \frac{\alpha G_F |V_{tb}V_{ts}^*|}{\sqrt{2} \pi}\right)^2\frac{ f_{B_s}^2 M_{B_s}^5\tau_{B_s}}{32 \pi (m_b+m_s)^2 } \left(1-\frac{4 m_\mu^2}{M_{B_s}^2}\right)^{1/2} \Big( \Big(1-\frac{4 m_\mu^2}{M_{B_s}^2}\Big)|C_S-C_S'|^2\nonumber \\
&+&|C_P-C_P'+2 \frac{m_\mu}{M^2_{B_s}}C_{10}|^2\Big) ~. \end{eqnarray}
The explicit expression for $C_{10}$ in  SM can be found for example in \cite{c10}. In view of the smallness 
of this contribution,  we would be interested in exploring the region of parameter space where the Higgs contribution significantly dominates over the contribution from $C_{10}$. It is thus useful to separate out the Higgs contribution $B_H$ alone to the above branching ratio and we define:

\be \label{bh}
B_H\equiv\left( \frac{\alpha G_F |V_{tb}V_{ts}^*|}{\sqrt{2} \pi}\right)^2\frac{ f_{B_s}^2 M_{B_s}^5\tau_{B_s}}{32 \pi (m_b+m_s)^2 } \left(1-\frac{4 m_\mu^2}{M_{B_s}^2}\right)^{1/2} \left( \left(1-\frac{4 m_\mu^2}{M_{B_s}^2} \right)|C_S-C_S'|^2+|C_P-C_P'|^2\right) ~. \ee
We however use the full equation, (\ref{bsh}) in our numerical study.

The  process $\bar{B}_d\rightarrow \bar{K}\mu^+\mu^-$ is studied in detail in \cite{bobeth1,bobeth2} using the QCD factorization approach which works for the low $q^2$ region. Restricting the dilepton invariant  (mass)$^2$ between the range
$1\text{GeV}^2<q^2<7\text{GeV}^2$, Bobeth {\it et al} derive \cite{bobeth2}
\begin{eqnarray}\label{brbdkmumu}
Br(\bar{B}_d \rightarrow \bar{K} \mu^+ \mu^-) &=& \Big(\frac{\tau_B^+}{1.64 \text{ps}}\Big) \Big( 1.91 + 0.02 (|\tilde{C_S}|^2 + |\tilde{C_P}|^2) - \frac{m_{\mu}}{GeV} \frac{Re(\tilde{C_P})}{2.92} \\ \nonumber
& & - \frac{m_{\mu}^2}{GeV^2} \Big(\frac{|\tilde{C_S}|^2}{5.98^2} + \frac{|\tilde{C_P}|^2}{10.36^2}\Big) + O(m_{\mu}^3)\Big) \times 10^{-7}
\end{eqnarray}
where $\tilde{C_S},\tilde{C_P}$ are given in terms of $C_{S,P},C'_{S,P}$ in eq.(\ref{cscp}) by
\begin{eqnarray}
\tilde{C_S} &=& C_S + C_S' ~, \nonumber\\
\tilde{C_P} &=& C_P + C_P'~.
\end{eqnarray}
\section{Constraining the FCNC couplings}
Among the processes mentioned above, the \bsbsbar transition is the most accurately measured and provide sensitive test of 
the FCNC couplings. In particular, the presence of these couplings in some cases can explain the additional CP violating 
phase in the $B_s\rightarrow J/\psi \phi $ decay.

The new physics contribution to \bsbsbar mixing is parametrized in terms of 
\begin{eqnarray} \label{cnp}
C_{B_s}   &=& | 1 + \kappa_s^H e^{2 i (\phi_s^H +\beta_s)} | ~, \nonumber\\
\phi_{B_S} &=& -\frac{1}{2} \text{Arg}(1 + \kappa_s^H e^{2 i (\phi_s^H + \beta_s)})~,
\end{eqnarray}
where $\kappa_s^H$ is given in our case by eq.(\ref{kapas}).
The 95\% allowed ranges of $C_{B_s}$ and $\phi_{B_s}$  given by UTfit collaboration are	\cite{Bonna}
\begin{eqnarray} \label{cnpvalue}
 C_{B_s} &=& [0.68 , 1.51] ~, \nonumber \\
\phi_{B_s} &=& [-30.5 , -9.9] \cup [-77.8 , -58.2]
\end{eqnarray}
We shall derive constraints on $F_{23},F_{32}$ based on the above values and look at its observable consequences for 
the processes $\bar{B}_s\rightarrow
\mu^+\mu^-,\bar{B}_d\rightarrow \bar{K}\mu^+\mu^-$. The derived constraints depend on the Higgs masses and mixing 
angles. But a simple and $F_{ij}$-independent correlations between $\kappa_s^H$ and the Higgs contribution $B_H$ to the branching ratio
for the process $\bar{B}_s\rightarrow
\mu^+\mu^-$ follows in the decoupling limit if it is assumed that the Higgs potential is the same as in the case of MSSM. We first derive
this relation. Then we give up these simplifying assumptions in the Higgs sector and explore the Higgs parameter space numerically and study the correlation between $\kappa_s^H$ and  the $\bar{B}_s\rightarrow
\mu^+\mu^-$ branching ratio. \\

The Higgs masses and mixing angle satisfy the following two relations \cite{hk, babukolda} if the scalar potential coincide with the 
MSSM.
\begin{eqnarray}\label{mssmrelation}
\langle\phi_F |\phi_F\rangle &=&0\,\nonumber \\
\cos^2(\alpha-\beta)&=&\frac{M_h^2(M_Z^2-M_h^2)}{M_A^2(M_H^2-M_h^2)}~.
\end{eqnarray}
The first relation leads to the following simple expression for $\kappa_s$:
\be \label{kapassimple}
\kappa_s^H e^{2 i \phi_s^{H}}=-\frac{4 \pi^2 P_2 F_{32} F_{23}^*}{B_{B_s}\eta_B S_0(x_t)G_F^2M_W^2 |V_{tb}^*V_{ts}|^2 M_A^2 } ~. \ee

Note that $$e^{2i\phi_s^H}=-\frac{F_{32}F_{23}^*}{|F_{32}F_{23}^*|}$$
directly probes the CP violating phase in the FCNC couplings and would depend on the model for quark masses under consideration.
\begin{itemize}
\item In models with Hermitian mass matrices, $\phi_s^H=Arg(F_{32})\pm\pi$ . This class of models can account for possible large CP violating phase $\phi_s$.
\item In contrast, the models with symmetric mass matrices, automatically imply $\phi_s^H=\pm \pi$. Thus even the presence of FCNC in these models does not lead
to large CP violation. Alternative source of CP violation can arise in these models if the Higgs sector violate CP. In this case, mixing between the scalar
and pseudo-scalar generate additional phase which can contribute to $\phi_s^H$. This scenario was studied in \cite{bhavik1} in a specific model with symmetric 
quark mass matrices.
\item $\phi_s^H$ is again given by the phase of $F_{32}$ in class (C) models satisfying $F_{32}=\frac{m_s}{m_b} F_{23}^*$. In particular, MSSM 
with MFV as well as the 2HDM of ref(\cite{bgl}) predict $F_{32}\sim V_{tb}^* V_{ts}$.  As a consequence, the Higgs generated phase $\phi_s^H$ coincide with 
the SM phase $\beta_s$ which is known to be small. Thus, these type of models will also need additional source, e.g. scalar-pseudoscalar mixing if
large $\phi_s$ is established.
\end{itemize}

The magnitude $\kappa_s^H$ of the Higgs contribution to the $B^0_s-\bar{B}^0_s$ mass difference relative to the SM contribution 
can be quite large for reasonable values of the unknown parameters.  Eq.(\ref{kapassimple}) implies that
\begin{equation} \label {kapasnum}
\kappa_s^H\approx 0.6 \left(\frac{F_{23}^*F_{32}}{10^{-6}}\right)\left(\frac{300{\mbox GeV}}{M_A}\right)^2 ~.\end{equation}
Consider various  model expectations: 
\begin{itemize}
\item If one uses the Cheng-Sher ansatz eq.(\ref{chengsher}) then $|F_{23}F_{32}|\approx {\cal O}(1) \frac{m_sm_b}{v^2}\approx  10^{-5}$. Eq.(\ref{kapasnum}) then gives
large contribution to $\kappa_s^H$.
\item Eq.(\ref{fcnc1}) gives the typical magnitude of FCNC in class of models discussed in section (2A).  In case of Hermitian textures with   $|F_{32}|=|F_{23}|\sim \frac{m_b}{v\cos\beta\sin\beta}| V_{L33}^*V_{L32}|$ we obtain 
$|F_{23}^*F_{32}|\sim 10^{-6}$ if $V_L\sim V$ leading to  a sizable value of 
$\kappa_s^H$ in this case also.
\item Models with $F_{32}^*=\frac{m_s}{m_b} F_{23}$ have additional suppression by $\frac{m_s}{m_b}$  compared to the previous estimates 
and one would need a light $A$ to obtain significant $\kappa_s^H$. There is also an additional suppression by loop factors in MSSM but the $F_{ij}$ can get enhanced by  $\tan\beta$. Typical magnitude of $F_{23}$ in MSSM is given by \cite{buras}
$$ F_{23}\approx \frac{g |V^*_{tb}V_{ts}| m_b\epsilon_Y }{\sqrt{2} M_W}\tan^2\beta~,$$
where $\epsilon_Y$ depends on the squark masses, the trilinear coupling $A_t$ and $\mu$. Taking the former two at TeV and $\mu\sim 300$ GeV, $\epsilon_Y\sim 0.002$ leading to
$F_{23}\sim 2  ~10^{-6} \tan^2 \beta$. Thus one can get significant effect only for very large $\tan\beta$
\end{itemize}

The expression for  $B_H$ gets simplified in the decoupling limit corresponding to $M_A^2\sim M_H^2\gg M_Z^2,M_h^2$ .
In this limit, $\alpha-\beta\rightarrow \frac{\pi}{2}$ from eq.(\ref{mssmrelation}) and the
couplings $C_{S,S',P,P'}$ satisfy
\be  \frac{C_S}{F_{23}}\approx \frac{C_S'}{F_{32}^*}\approx -\frac{C_P}{F_{23}}\approx \frac{C_P'}{F_{32}^*}  \approx \frac{\sqrt{2}\pi m_\mu}{\alpha G_F V_{tb}V_{ts}^*}\frac{\sin\beta}{2 v \cos\beta M_A^2}~.\ee
Because of this, the $B_H$ in eq.(\ref{bh}) reduces to
\begin{equation} \label{bshsimple}
B_H=\frac{f_{B_s}^2M_{B_s}^5\tau_{Bs}}{128 \pi (m_b+m_s)^2} \left(\frac{m_\mu^2}{ v^2}\right)\frac{\tan^2\beta}{M_A^4}\left(1-\frac{4 m_\mu^2}{M_{B_s}^2}\right)^{1/2} \left( \left(1-\frac{4 m_\mu^2}{M_{B_s}^2}\right)|F_{23}-F_{32}^*|^2+|F_{23}+F_{32}^*|^2\right) ~. \end{equation}

The above equation allows us to derive simple correlation between $\kappa_s^H$
and $B_H$. Combining eqs.(\ref{kapassimple}) and (\ref{bshsimple}) we find  
\begin{eqnarray} \label{relation}
B_H&\approx& \frac{4 b \kappa_s^H \tan^2\beta}{\kappa M_A^2}\approx 2.2 ~10^{-8}\kappa_s^H \left(\frac{\tan\beta}{50}\right)^2\left(\frac{300 {\mbox GeV}}{M_A}\right)^2 ~~~~~~~~~~~~{\rm (Models (A) \& (B))}~,\nonumber\\  
&\approx& \frac{2 b \kappa_s^H \tan^2\beta}{\kappa M_A^2}\frac{m_b}{m_s}\approx 1.7~~ 10^{-8}\kappa_s^H \left(\frac{\tan\beta}{10}\right)^2\left(\frac{300 {\mbox GeV}}{M_A}\right)^2 ~~~~~~~~~~~~{\rm (Models (C)}~,  \end{eqnarray}
where 
$$b\equiv \frac{f_{B_s}^2 M_{B_s}^5\tau_{B_s}}{128 \pi(m_b+m_s)^2}\left(\frac{m_\mu}{v}\right)^2\approx 1.1 ~ 10^4 {\mbox GeV}^4~,$$
$$ \kappa\equiv \frac{4 \pi^2}{B_{B_s}\eta_BS_0(x_t)G_F^2 M_W^2 |V_{tb}^*V_{ts}|^2}\approx 2.2 ~ 10^{10} {\mbox GeV}^2 ~.$$
These correlations are independent of the magnitude and phases of the FCNC couplings and therefore test the assumption of (1) the presence of FCNC
and (2) the MSSM structure in the Higgs potential independent of the detailed structures of the quark mass matrices. These correlations also show that the FCNC would lead to sizable $B_H$ 
provided it gives significant correction to $\kappa_s^H$ also.
\begin{figure}[t]
\centering
\includegraphics{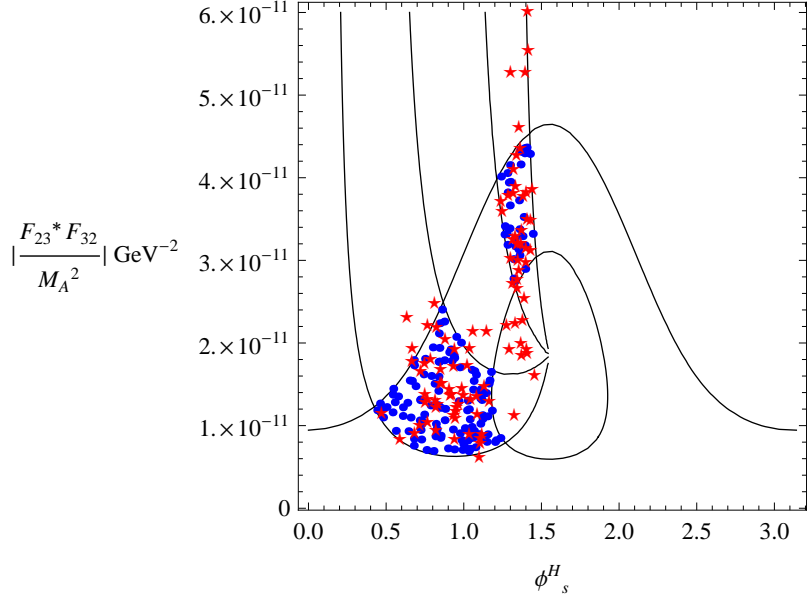}
\caption{The region in $|\frac{F_{32} F^*_{23}}{M^2_A}| - \phi^H_s$ allowed by the UTFit constraints on \bsbsbar mixing. The solid lines and dots describe the region allowed under the assumption of the same  Higgs potential as in MSSM. The stars correspond to assuming general Higgs sector and varying parameters as explained in the text.}
\label{fig:1}
\end{figure}

Let us now turn to the numerical analysis. If we assume the MSSM like Higgs structure
then the allowed ranges of $\phi_{B_s}$ and $C_{B_s}$ given in (\ref{cnpvalue}) determines the magnitude and phase of $F_{32}F_{23}^*$, see. eq.(\ref{kapassimple}).  The allowed region in $|\frac{F_{32}F_{23}^*}{M_A^2}|$-$\phi_s^H$ plane is shown in Fig.(1).
No specific assumption is made on the nature of the FCNC couplings. Therefore fig.(1) represents generic constraints on these
couplings in all the 2HDM with tree level FCNC. The allowed values of $|F_{32}F_{23}^*|$ typically lie in the region $(1-5)\times 10^{-11} M_A^2 ~{\rm GeV}^{-2}$ with a strong correlation between its magnitude and phase.
A generic 2HDM need not follow the MSSM structure and the decoupling would also correspond to only a part of the available parameter space.
We study departures from these assumptions  numerically as follows. We 
randomly vary the Higgs masses $M_h,M_{H},M_A$  between the range $100-500$ GeV keeping $M_{h}\leq M_H$. The mixing angles $\alpha,\beta$ are
varied in their full range. From every set of these input parameters we allow those which give $C_{B_s},\phi_{B_s}$ in the range in eq.(\ref{cnpvalue}) and the $Br( \bar{B}_s\rightarrow \mu^+\mu^-) $ below the limit in eq.(\ref{bsmummu}). In this random analysis we distinguish two cases.  One in which the MSSM relation eq.(\ref{mssmrelation}) remains true. These cases are shown as dots in our figure while the more general case without that assumption is shown as $\star$.

\begin{figure}[t]
\centering
\includegraphics{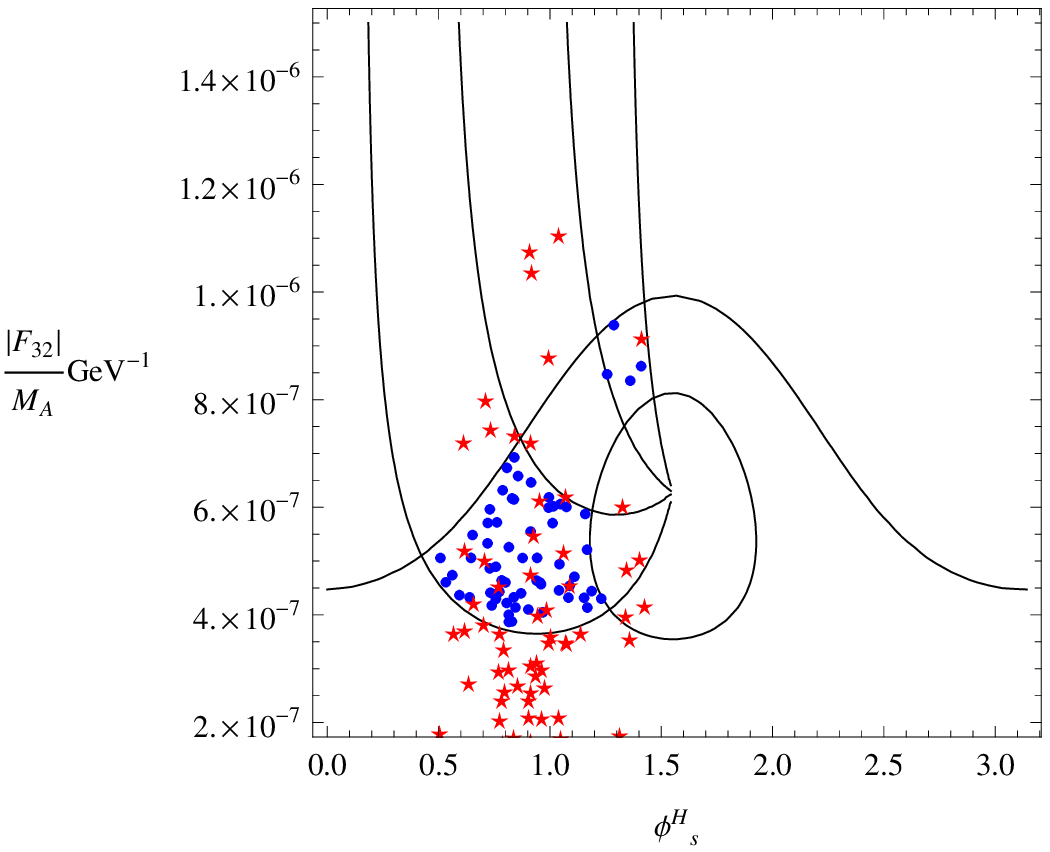}
\label{fig:2}
\caption{The region in $|\frac{F_{32}}{M_A}| - \phi^H_s$ allowed by the \bsbsbar mixing constraints in eq.(\ref{cnpvalue}) in class of models satisfying $F_{32} =\frac{m_s}{m_b} F_{23}^*$. Other details are as in Fig.(1)} 
\end{figure}

Fig(2) shows the allowed region in the $\frac{|F_{32}|}{M_A}$-$\phi_s^H$ plane in classes of models which satisfy the constraints $F_{32}=\frac{m_s}{m_b}F_{23}^*$.
One obtains the constraint $|F_{32}|\lesssim 1.2\times 10^{-6} \frac{M_A}{{\rm GeV}}$. This is to be compared with typical MSSM value
$1.6~~ 10^{-6} \tan^2\beta$. Thus one would need $\tan^2\beta\approx \frac{M_A}{{\rm GeV}}$ to account for the magnitude $C_{B_s}$. 
If $F_{23}$ is given by eq.(\ref{fcnc1}) then $F_{32}\approx 3\times 10^{-5}\frac{1}{\sin\beta\cos\beta}\frac{.05}{|V_{L23}V_{L33}|}$. Thus, in this class of models one would need $|V_{L23}|$ somewhat smaller than $|V_{cb}|\sim 0.05$. In contrast to MSSM, large values of $\tan\beta$ are disfavored by the \bsbsbar mixing constraint in this class of models. 

\begin{figure}[t]
\centering
\includegraphics{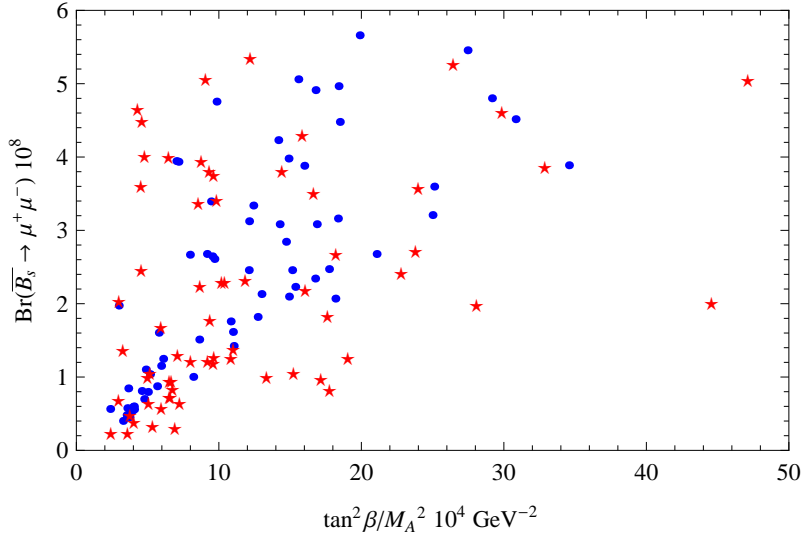}
\caption{Variations for the  branching ratio of the process $\bar{B}_s \rightarrow \mu^+ \mu^-$ with respect to $\tan^2\beta / M^2_A$ after incorporating the \bsbsbar constraints in model with $F_{32} = \frac{m_s}{m_b} F_{23}^*$. The dots and stars are defined as in Fig.(1) } 
\label{fig:3}
\end{figure} 

Fig.(3) shows the allowed values of the branching ratio for $\bar{B}_s\rightarrow \mu^+\mu^-$ obtained under the assumption
$F_{32}=\frac{m_s}{m_b}F_{23}^*$ after imposing the UTfit constraints.  It is possible to obtain relatively large branching ratios  even for moderate values of $\tan\beta$ if $M_A$ is light $\sim 100~$ GeV.

\begin{figure}[t]
 \includegraphics{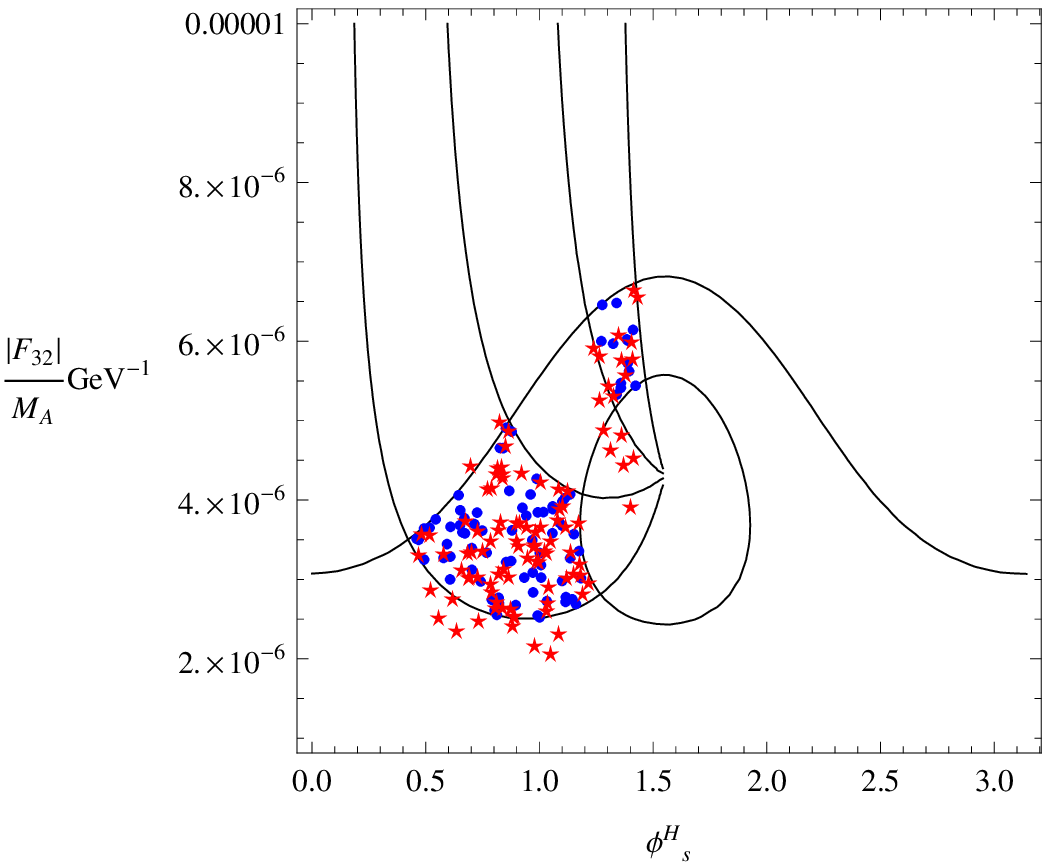}
\caption{The region in $|\frac{F_{32}}{M_A}| - \phi^H_s$ allowed by the \bsbsbar mixing constraints in eq.(\ref{cnpvalue}) in class of models satisfying $F_{32} = F_{23}^*$. Other details are as in Fig.(1)} 
\label{fig:4}
\end{figure}

Fig.(4)  represents the corresponding constraints in class of models with Hermitian structure $F_{23}=F_{32}^*$. The 
required values for $F_{32}$ are now $(2-6)\times 10^{-6} M_A$ . But once again, one could obtain measurable rate for the dimuonic $B_s$ decay even with 
moderate value of $\tan\beta$ as shown in Fig.(5).

\begin{figure}[t]
\centering
\includegraphics{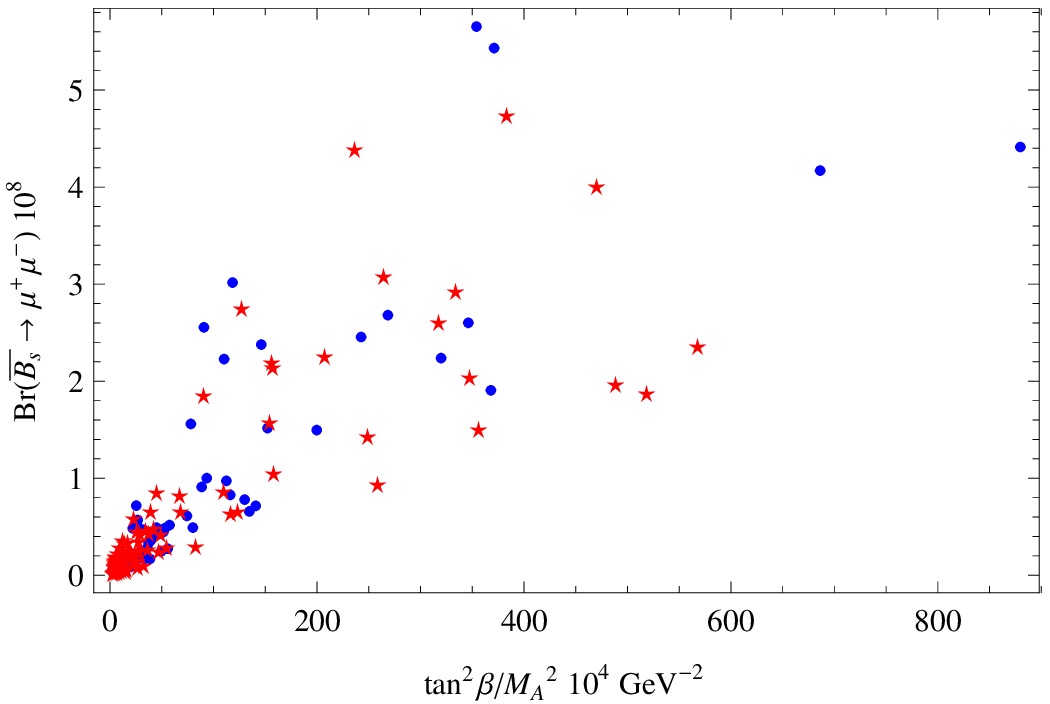}
\caption{Variations for the  branching ratio of the process $\bar{B}_s \rightarrow \mu^+ \mu^-$ with respect to $\tan^2\beta / M^2_A$ after incorporating the \bsbsbar constraints in model with $F_{32} = F_{23}^*$. The dots and stars are defined as in Fig.(1)}
\label{fig:5}
\end{figure}

\begin{figure}[t]
\centering
\includegraphics{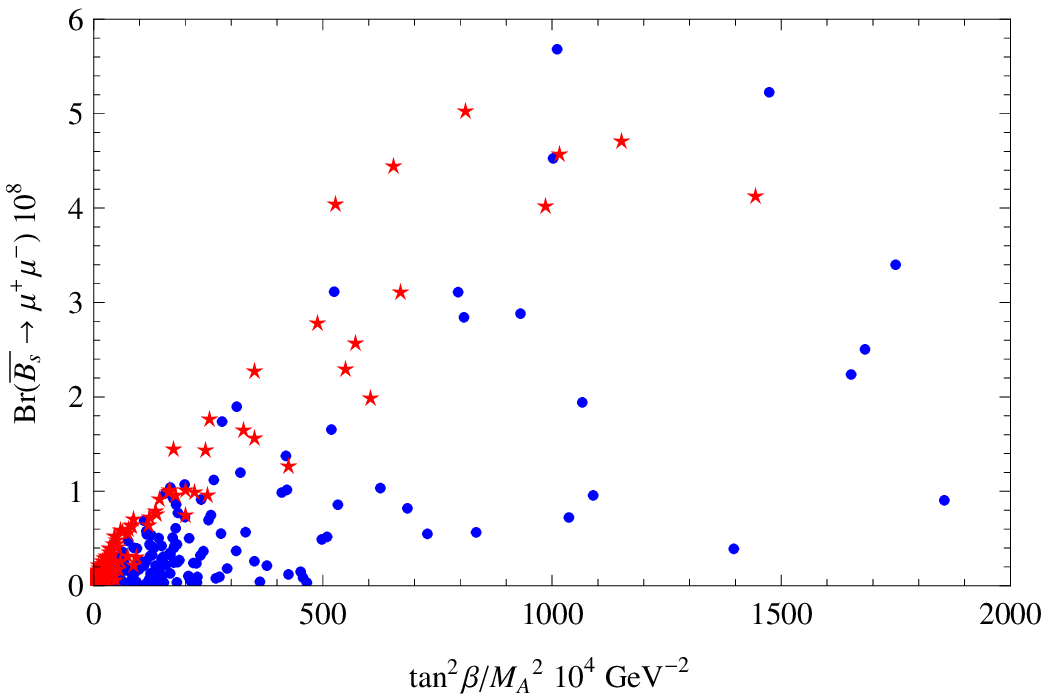}
\caption{Variations for the  branching ratio of the process $\bar{B}_s \rightarrow \mu^+ \mu^-$ with respect to $\tan^2\beta / M^2_A$ after incorporating the \bsbsbar constraints in model with $F_{32} = F_{23}$. The dots and stars are defined as in Fig.(1)}
\end{figure}

Fig.(6) displays the allowed  values of $\bar{B}_s\rightarrow \mu^+\mu^-$ in the case $F_{23}=F_{32}$.  It is seen
 that one needs relatively large $\tan\beta$ typically $\tan^2\beta/M_A^2 \approx 10^{-2} GeV^{-2}$ in order to obtain
a branching ratio larger than $10^{-8}$. As already mentioned, this case also predicts vanishing Higgs induced phase
if the Higgs sector is CP conserving.

While $\bar{B}_s\rightarrow \mu^+\mu^-$ can receive significant contribution from the FCNC, the same is not the case
with the semi leptonic process $\bar{B}_d\rightarrow \bar{K}\mu^+\mu^-$. The FCNC induced contribution to this process can be
qualitatively different than the 2HDM model based on the NFC. For example, if $F_{23}=F_{32}^*$ then eq.(\ref{bh}) and eq.(\ref{brbdkmumu}) together imply that only
the scalar Higgses contribute to $\bar{B}_d\rightarrow \bar{K}\mu^+\mu^-$ while  $\bar{B}_s\rightarrow \mu^+\mu^-$ gets contribution from
the pseudoscalar Higgs. Thus these processes are uncorrelated if the corresponding  Higgs masses are not correlated. This is to be compared
with the standard 2HDM or the MSSM where definite correlations between these processes have been pointed out \cite{amol}. At the quantitative level, we find
numerically that after imposition of the $B_s^0-\bar{B}_s^0$ mixing, the allowed numerical values of the couplings $\tilde{C}_{S,P}$ in all cases are such that the Higgs contribution to the branching ratio of $\bar{B}_d\rightarrow \bar{K}\mu^+\mu^-$amounts to at most few percent of the SM contribution . This is much smaller than the  theoretical uncertainties. Therefore detecting Higgs
effects in this branching ratio would need considerable reduction in theoretical errors.  However one can conclude that if a significant new physics  contribution to the branching ratio of this process is detected, it cannot be due to the presence of the Higgs induced FCNC.

\section{Conclusion}

$b\rightarrow s$ transition is known to be a good probe of physics beyond standard model. We have looked at the possibility of 
using this transition to test the Higgs induced FCNC assuming that the neutral Higgs provides the dominant contribution. In this case, several processes get described in terms of two complex parameters $F_{23}$ and $F_{32}$ and the Higgs mass parameters through equation(\ref{fcncint}). Phenomenological analysis in  many of the earlier works \cite{type3phen}  used the specific form for $F_{23}$ and $F_{32}$
motivated by the Cheng-Sher ansatz and  often considered them to be real. We have tried to develop model-independent constraints
on these parameters. In particular, as shown here, the phases of the FCNC  couplings can
play an important role and may provide the large CP violating phase that may be needed to explain the CDF and $D0$ results on CP violation. \\
We discussed phenomenology of three broad classes of theories with FCNC satisfying the relations (1) $F_{23}=F_{32}^*$ (2) $F_{23}=F_{32}$ and (3) $F_{32}=\frac{m_s}{m_b}F_{23}^*$. We discussed several textures of the Yukawa couplings giving
rise to these relations. In particular, MSSM
and  2HDM with NFC provide examples of (3).  We showed that the case (2) cannot account for large CP violating phase if the Higgs sector is CP conserving. The same applies to MSSM and the particularly predictive model of \cite{bgl}. 
Our numerical analysis shows that one typically needs $F_{32}\sim (10^{-6}-10^{-7}) M_A$ GeV$^{-1}$. As discussed here such values 
can arise within the textures discussed in section (II).  \\
Using the available information on the \bsbsbar mixing we have worked out expectations for the leptonic branching ratio $\bar{B}_s\rightarrow \mu^+\mu^-$. It is found that the former constraints do allow measurable values for this branching ratio but the range for $\frac{\tan^2\beta}{M_A^2}$ required in these cases are different as seen from Figs. (3,5,6). In contrast, the Higgs contribution to the branching ratio of process $\bar{B}_d\rightarrow \bar{K}\mu^+\mu^-$ is constrained to be close to or smaller than  the SM value in all these models. Thus any significant deviation
in this branching ratio compared to the SM prediction will rule out all the models with FCNC in one shot under the assumption that these models 
are the only source of new physics in the \bsbsbar mixing.\\ \\
\noindent{\bf Acknowledgements:} We thank Namit Mahajan for several useful discussions.\\

\end{document}